\documentclass[
    aps,
    twocolumn,
    superscriptaddress,
    preprintnumbers
]{revtex4}

\usepackage{graphicx}  
\usepackage{subfigure}
\usepackage{multirow}

\usepackage{fancyhdr}
\usepackage{longtable}
\usepackage{parskip}
\usepackage[T1]{fontenc}
\usepackage{dcolumn}   

\usepackage{bm}        
\usepackage{amsfonts}  
\usepackage{amsmath}   
\usepackage{amssymb}   
\usepackage{upgreek}
\usepackage{hyperref}
\usepackage{xr-hyper}

\newcommand{\pwisein}{\left\{ \begin{array}{ll}}
\newcommand{\pwiseout}{\end{array}\right.}

\begin{document}


\title{Hysteresis without coexistence: disorder-rounded first-order transitions in a van der Waals magnet}

\author{Xiaoyu Guo}
\affiliation{William H. Miller III Department of Physics and Astronomy, The Johns Hopkins University, Baltimore, Maryland 21218, USA}

\author{Abby N. Neill}
\affiliation{Department of Chemistry,
The Johns Hopkins University, Baltimore, Maryland 21218, USA}

\author{Christopher M. Pasco}
\affiliation{Department of Chemistry,
The Johns Hopkins University, Baltimore, Maryland 21218, USA}

\author{Tyrel M. McQueen}
\affiliation{William H. Miller III Department of Physics and Astronomy, The Johns Hopkins University, Baltimore, Maryland 21218, USA}
\affiliation{Department of Chemistry,
The Johns Hopkins University, Baltimore, Maryland 21218, USA}
\affiliation{Department of Materials Science and Engineering,
The Johns Hopkins University, Baltimore, Maryland 21218, USA}

\author{N. P. Armitage}
\email[Corresponding author: ]{npa@jhu.edu}
\affiliation{William H. Miller III Department of Physics and Astronomy, The Johns Hopkins University, Baltimore, Maryland 21218, USA}

\date{\today}

\begin{abstract}  

Quenched disorder can profoundly modify phase transitions. In low-dimensional systems, theory predicts that even weak quenched disorder can round the thermodynamic discontinuities associated with a first-order phase transition. 
Here, we employ time-domain terahertz spectroscopy to investigate the quasi-two-dimensional trimerized kagome van der Waals magnet family Nb$_3$Cl$_{8-x}$Br$_x$ ($x=0$, 1 and 8). We observe the emergence of an additional phonon branch upon Br substitution, whose spectral weight increases and frequency softens with increasing Br concentration. The temperature evolution of the phonon frequencies reveals a clean first-order transition in Nb$_3$Cl$_8$ characterized by macroscopic phase coexistence  and thermal hysteresis. In contrast, the transition in the substitutionally disordered compound Nb$_3$Cl$_7$Br retains its hysteresis while exhibiting a substantially broadened transition with no resolvable macroscopic phase coexistence.  These observations reveal disorder-induced fragmentation of the transition into locally favored domains instead of well-defined bulk phases separated by stable phase boundaries.  The behavior is consistent with the Imry-Wortis and the Aizenman-Wehr scenarios for the effect of quenched disorder in low-dimensional systems, which destabilizes macroscopic phase coexistence and rounds the thermodynamic discontinuities associated with first-order transitions.  Thermal hysteresis persists in the disordered compound despite the lack of resolvable coexistence, indicating that the two features often treated as a single hallmark of first-order character arise distinctly and can be separated by disorder.  Moreover, our results establish Nb$_3$Cl$_{8-x}$Br$_x$ as a promising platform for investigating the effects of disorder on first-order transitions in low-dimensional systems.

\end{abstract}

\maketitle 

\section{Introduction}

Disorder can affect phase transitions in fundamental ways. In continuous (second-order) transitions, the effect of quenched disorder has long been understood through the Harris criterion~\cite{harris_effect_1974}, which establishes the condition where disorder modifies the universal critical behavior. In contrast, the effects of disorder on first-order phase transitions remain less understood~\cite{cox_effect_1988}. Owing to metastability and finite energy barriers separating competing phases, first-order transitions are often accompanied by thermal hysteresis and phase coexistence.  These two signatures are frequently treated interchangeably, but they reflect different physics: coexistence results from two phases of equal free energy density, whereas hysteresis reflects a kinetic barrier to switching between them -- neither necessarily implies the other.

Theoretical studies have shown that quenched disorder can destabilize macroscopic phase coexistence and fragment the transition into spatially separated domains, while maintaining hysteresis~\cite{imry_influence_1979}. In sufficiently low-dimensional systems, disorder is further predicted to round or even smear the discontinuity associated with the transition altogether~\cite{aizenman_rounding_1989}. Despite the prevalence of first-order transitions in condensed matter systems in the presence of disorder, such as liquid crystals~\cite{iannacchione_review_2004,ramazanoglu_first-order_2004}, alloys~\cite{roy_first_2004}, magnets~\cite{otero-leal_quenched_2007} and metal-insulator transitions~\cite{tan_unraveling_2012, vidas_imaging_2018, gati_effects_2018, mattoni_striped_2016, nathwani_observation_2026}, experimental investigations of disorder-mediated first-order transitions in low dimensions remain scarce~\cite{cervera_mastering_2024, nathwani_observation_2026}. This is largely because few model systems simultaneously exhibit a clean first-order transition, permit controlled tuning of disorder, and offer experimentally accessible probes of the transition.

The trimerized kagome van der Waals (vdW) materials Nb$_3$X$_8$ (X = Cl, Br, and I) provide an attractive platform for investigating these questions. Nb$_3$X$_8$ consists of weakly coupled layers of Nb trimers arranged on a breathing kagome lattice, giving rise to a quasi-two-dimensional (quasi-2D) crystal structure. Previous studies have shown that varying the halogen species systematically modifies the lattice and electronic structures while preserving the underlying trimerized kagome structure~\cite{pasco_tunable_2019, date_momentum-resolved_2025, regmi_observation_2023, regmi_spectroscopic_2022, sun_observation_2022, mortazavi_first-principles_2022}. This provides an opportunity to introduce controlled disorder without changing the underlying crystal framework. Theoretical calculations further suggest that the electronic ground state evolves from a strongly correlated Mott-insulating regime in Nb$_3$Cl$_8$ toward a more weakly correlated band-insulating regime in Nb$_3$I$_8$~\cite{aretz_strong_2025}. Importantly, the Nb$_3$X$_8$ family exhibits  coupled structural and magnetic transitions with pronounced first-order characteristics into a low-temperature singlet state that is accompanied by a rearrangement of the stacking sequence between vdW layers~\cite{haraguchi_magneticnonmagnetic_2017, sheckelton_rearrangement_2017, pasco_tunable_2019, kim_terahertz_2023}. The transition temperature increases systematically from Nb$_3$Cl$_8$ to Nb$_3$Br$_8$ with increasing Br substitution~\cite{pasco_tunable_2019}, whereas the transition in Nb$_3$I$_8$ may occur well above room temperature~\cite{kim_terahertz_2023, aretz_strong_2025}. Combined with their quasi-2D nature and broad chemical tunability, these characteristics establish Nb$_3$X$_8$ as a promising platform for studying the interplay between disorder and first-order phase transitions.

\begin{figure*}[t]
    \centering
    \includegraphics[
        trim=0cm 0cm 0cm 0cm,
        clip,
        width=\textwidth
    ]{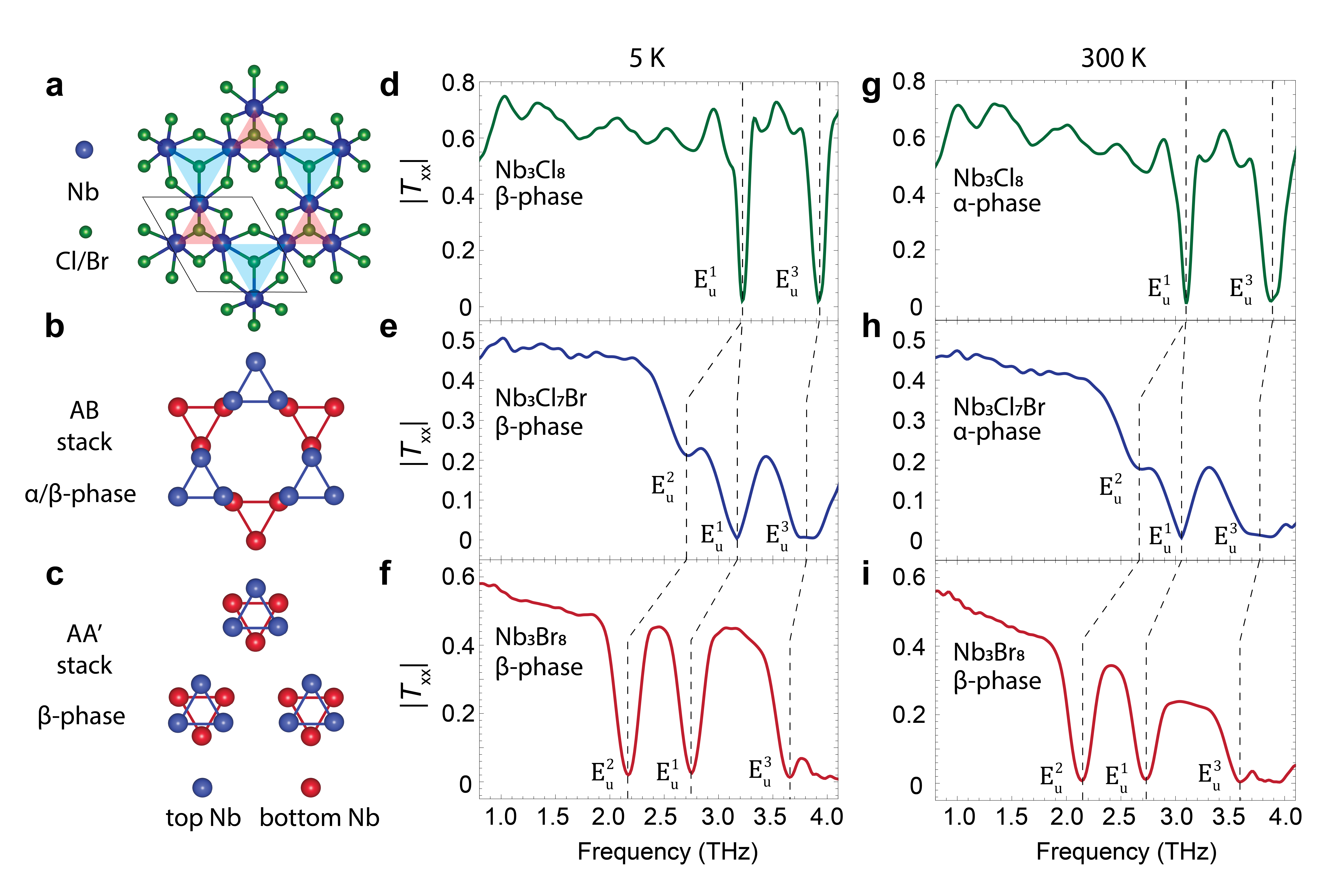}
    \caption{\textbf{Atomic structure of Nb$_3$X$_8$ and evolutions of IR-active phonons with different halogen compositions.} (a) Atomic structure of monolayer Nb$_3$X$_8$. The breathing kagome lattice is shaded in blue and red and the unit cell is indicated by the parallelogram. (b-c) The AB (b) and AA$^\prime$ (c) stacking of the Nb atoms in adjacent vdW layers. $\alpha$-phase Nb$_3$X$_8$ only has AB stacking while $\beta$-phase has alternating AB and AA$^\prime$ stackings. (d-i) The amplitude of transmission coefficient $T_{xx}$ from Nb$_3$Cl$_8$, Nb$_3$Cl$_7$Br and Nb$_3$Br$_8$ at 5 K (d-f) and at 300 K (g-i). The corresponding phonon modes are labeled.}
    \label{fig:highlowTSpectra}
\end{figure*}

Here, we investigate the evolution of first-order phase transitions in isostructural Nb$_3$X$_8$~\cite{aretz_strong_2025} using time-domain terahertz spectroscopy (TDTS) on Nb$_3$Cl$_8$, Nb$_3$Cl$_7$Br, and Nb$_3$Br$_8$. We find that halogen substitution introduces an additional phonon branch whose spectral weight gradually increases and frequency softens with increasing Br concentration. Several phonon modes exhibit pronounced frequency shifts across the coupled structural and magnetic transition, providing a sensitive probe to the transition. In the unsubstituted compound Nb$_3$Cl$_8$, distinct phonon branches associated with the high- and low-temperature phases coexist over a finite temperature interval, evidencing a macroscopic first-order transition. In contrast, the substitutionally disordered compound Nb$_3$Cl$_7$Br exhibits a continuous evolution of phonon frequencies with no resolvable phase coexistence but retaining hysteresis. This is indicative of a disorder-broadened first-order transition. In Nb$_3$Br$_8$, no phase transition has been captured by TDTS up to room temperature, consistent with the previously reported transition temperature of 382 K~\cite{pasco_tunable_2019}. Terahertz (THz) polarimetry measurements under magnetic fields further confirm that Br disorder preserves the nonmagnetic ground state.  These results establish Nb$_3$X$_8$ as a model quasi-2D platform for investigating disorder effects on the evolution of first-order phase transitions.

\begin{table*}[t]
    \centering
    \caption{$E_u$ phonon energies at 5 K and 300 K in Nb$_3$Cl$_{8-x}$Br$_x$ (unit: THz)}
    \label{tab:Eu_modes}
    \renewcommand{\arraystretch}{1.4}
    \begin{tabular}{|c|c|c|c|c|c|c|}
    \hline
    \makebox[2.3cm][c]{}
     & \makebox[2.3cm][c]{$E_u^1$ (5 K)}
     & \makebox[2.3cm][c]{$E_u^1$ (300 K)}
     & \makebox[2.3cm][c]{$E_u^2$ (5 K)}
     & \makebox[2.3cm][c]{$E_u^2$ (300 K)}
     & \makebox[2.3cm][c]{$E_u^3$ (5 K)}
     & \makebox[2.3cm][c]{$E_u^3$ (300 K)} \\
    \hline
    character
     & \multicolumn{2}{c|}{softened across phase transition}
     & \multicolumn{2}{c|}{activated by Br substitution}
     & \multicolumn{2}{c|}{insensitive to phase transition} \\
    \hline
    Nb$_3$Cl$_8$   & 3.22 & 3.11 & N/A  & N/A  & 3.92 & 3.90 \\
    Nb$_3$Cl$_7$Br & 3.15 & 3.03 & 2.72 & 2.65 & 3.84 & 3.82 \\
    Nb$_3$Br$_8$   & 2.75 & 2.73 & 2.17 & 2.13 & 3.56 & 3.60 \\
    \hline
    \end{tabular}
\end{table*}

Within each layer of Nb$_3$X$_8$, Nb atoms form a distorted breathing kagome lattice (Fig.~\textbf{\ref{fig:highlowTSpectra}a}), in which groups of three Nb atoms cluster into trimers, giving rise to alternating Nb-Nb bond lengths. Two stacking configurations are possible between adjacent layers. In the AB stacking configuration, Nb trimers in neighboring layers are arranged in a staggered fashion (Fig.~\textbf{\ref{fig:highlowTSpectra}b}), whereas in the AA$^\prime$ configuration, the trimers are aligned directly above one another (Fig.~\textbf{\ref{fig:highlowTSpectra}c}). In both cases, the orientation of the trimers rotates by 180$^\circ$ between adjacent layers along the crystallographic $c$ axis~\cite{habermehl_triniobiumoctabromide_2010, haraguchi_magneticnonmagnetic_2017, sheckelton_rearrangement_2017}.  Both Nb$_3$Cl$_8$ and Nb$_3$Br$_8$ undergo coupled structural and magnetic phase transitions. The transition temperature is approximately 90 K in Nb$_3$Cl$_8$ and about 382 K in Nb$_3$Br$_8$~\cite{pasco_tunable_2019}. In the high-temperature $\alpha$-phase, the structure consists of AB-stacked layers and adopts the $P\bar{3}m1$ space group with a two-layer unit cell. Upon cooling into the low-temperature $\beta$-phase, an alternating sequence of AB and AA$^\prime$stackings develops, enlarging the unit cell to six layers. The $\beta$-phase of Nb$_3$Br$_8$~\cite{habermehl_triniobiumoctabromide_2010} has been reported to crystallize in the $R\bar{3}m$ space group. In contrast, the low-temperature crystal symmetry of Nb$_3$Cl$_8$ remains under debate, with several space groups having been proposed, including $R3$~\cite{haraguchi_magneticnonmagnetic_2017}, $R\bar{3}m$~\cite{kim_terahertz_2023}, and $C2/m$~\cite{sheckelton_rearrangement_2017}.  In all cases the low-temperature phases are accompanied by an almost total loss of magnetization, indicating the formation of a singlet ground state~\cite{pasco_tunable_2019}.

\section{Experiment}

TDTS measurements were performed using a custom-built spectrometer equipped with fiber-coupled emitter and detector modules from the Toptica TeraFlash Pro system. Four off-axis parabolic mirrors arranged in an 8$f$ geometry provide a frequency-independent focus at the sample position. The transmitted electric fields through the sample, $\tilde{E}_{\mathrm{Samp}}(t)$, and through a reference aperture, $\tilde{E}_{\mathrm{Ref}}(t)$, were measured in the time domain. After Fourier transformation, the complex transmission coefficient

\begin{equation}
\tilde{T}(\omega)=\frac{\tilde{E}_{\mathrm{Samp}}(\omega)}
{\tilde{E}_{\mathrm{Ref}}(\omega)},
\end{equation}

\noindent was used together with the Fresnel equations to extract the complex refractive index $\tilde{n}$ according to~\cite{bilbro_fluctuations_nodate}

\begin{equation}
\tilde{T}(\omega)=\frac{4\tilde{n}}{(\tilde{n}+1)^2}
\exp\left[i\frac{\omega L}{c}(\tilde{n}-1)\right],
\end{equation}

\noindent where $L$ is the sample thickness.  $\tilde{n}$ was extracted by Newton$^\prime$s method implemented by the FindRoot function in Mathematica.  The complex optical conductivity, $\tilde{\sigma}=\sigma_1+i\sigma_2$, is related to the complex refractive index, $\tilde{n}=n+ik$, through~\cite{dressel_electrodynamics_2002}

\begin{equation}
\sigma_1(\omega)=2\epsilon_0\omega nk,
\end{equation}

\begin{equation}
\sigma_2(\omega)=
-\epsilon_0\omega\left(n^2-k^2-1\right).
\end{equation}

The incident THz pulse is linearly polarized and oriented away from the principal crystal axes. Two electro-optic detectors are employed to simultaneously measure the transmitted THz field components parallel and perpendicular to the incident polarization, yielding the complex transmission matrix elements $T_{xx}$ and $T_{xy}$.  Crystals analyzed in this work were grown following the chemical vapor transport procedure outlined in Pasco, et al.~\cite{pasco_tunable_2019}.  Experiments were performed in transmission on the crystals. The measured Nb$_3$Cl$_8$, Nb$_3$Cl$_7$Br and Nb$_3$Br$_8$ crystals have thicknesses 80 $\upmu$m, 120 $\upmu$m and 280 $\upmu$m, respectively.


\begin{figure*}[t]
    \centering
    \includegraphics[
        trim=0cm 0cm 0cm 0cm,
        clip,
        width=\textwidth
    ]{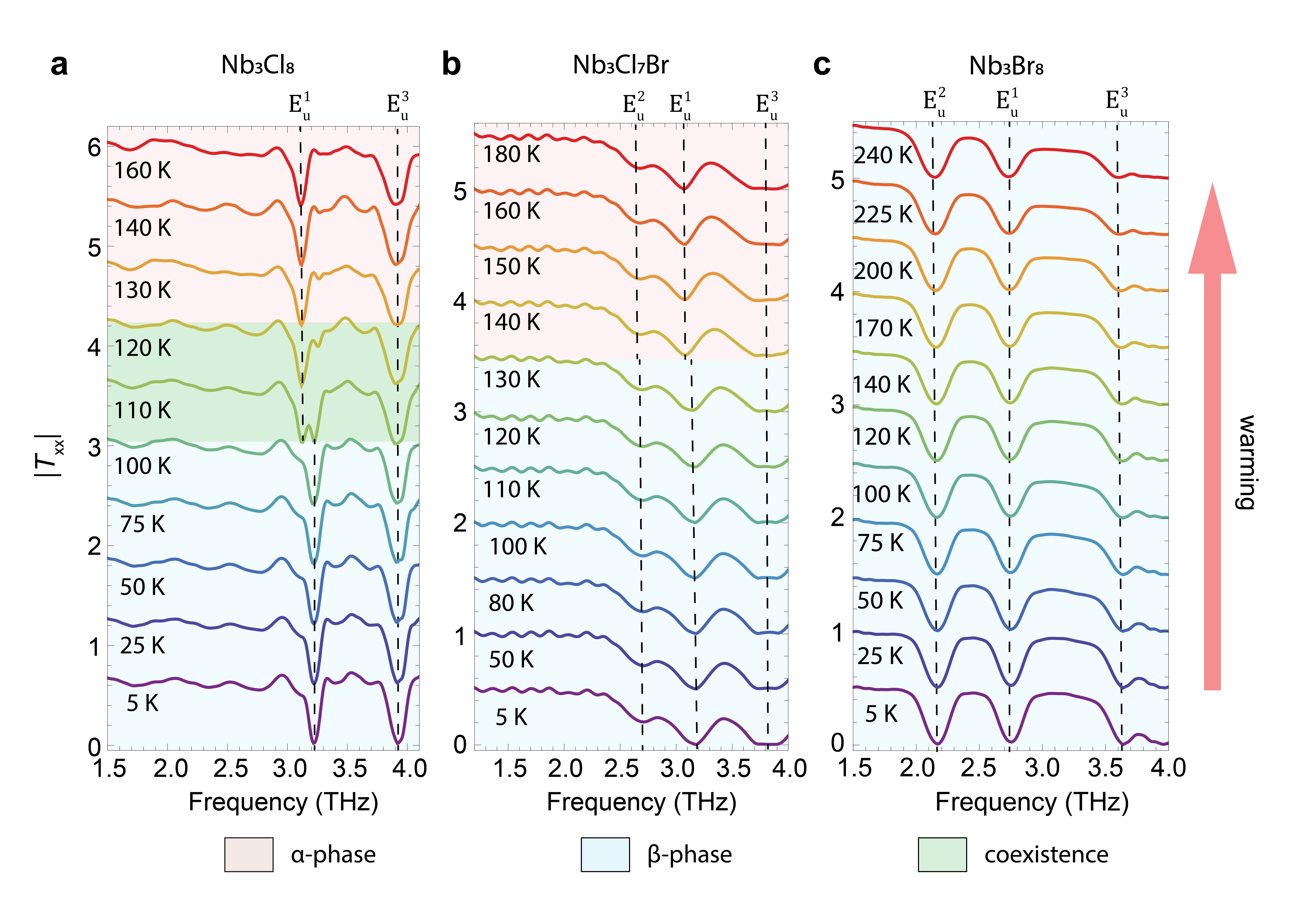}
    \caption{\textbf{$T$-dependent transmission spectra capture phase transitions and phase inhomogeneity.}
    (a-c) The magnitude of the transmission coefficient $T_{xx}$ as a function of temperature in Nb$_3$Cl$_8$ (a), Nb$_3$Cl$_7$Br (b) and Nb$_3$Br$_8$ (c). Measurements were taken during warming. The temperature ranges corresponding to the $\alpha$- and $\beta$-phases, and their coexistence are shaded in red, blue and green, respectively.
    }
    \label{fig:TdepSpectra}
\end{figure*}

\section{Results}

The low-energy lattice excitations of Nb$_3$Cl$_{8-x}$Br$_x$ ($x=0$, 1, and 8) measured by TDTS evolve systematically with Br concentration and provide sensitive signatures of the phase transition. Figures~\textbf{\ref{fig:highlowTSpectra}d-i} present a side-by-side comparison of the THz transmission spectra $|T_{xx}|$ for Nb$_3$Cl$_8$, Nb$_3$Cl$_7$Br, and Nb$_3$Br$_8$ at 5 K and 300 K, with the incident and detected THz polarizations parallel to each other. At 5 K (Figs.~\textbf{\ref{fig:highlowTSpectra}d-f}), all three compounds are in the low-temperature $\beta$-phase. In Nb$_3$Cl$_8$, two pronounced phonon peaks are observed at 3.22 THz and 3.92 THz. Based on the inversion symmetry and threefold rotational symmetry of the crystal, which are further confirmed by THz polarimetry measurements presented below, these infrared-active modes are assigned to doubly degenerate odd-parity $E_u$ phonons, denoted $E_u^1$ and $E_u^3$, respectively. A magnetic origin of these modes can be excluded, as their frequencies exhibit no measurable magnetic-field dependence~\cite{kim_terahertz_2023}.

In Nb$_3$Cl$_7$Br, where one of the eight halogen sites is occupied by Br, an additional weak absorption peak emerges at 2.72 THz. We assign this mode to a new phonon branch, denoted $E_u^2$. In Nb$_3$Br$_8$, where all halogen sites are occupied by Br, the spectral weight of $E_u^2$ increases and its frequency softens from 2.72 THz to 2.17 THz. The frequencies of $E_u^1$ and $E_u^3$ also decrease progressively with increasing Br concentration. The systematic softening of all three phonon branches with increasing Br content is qualitatively consistent with the larger halogen mass and the accompanying modification of interatomic force constants.

We next examine the $|T_{xx}|$ spectra at 300 K (Figs.~\textbf{\ref{fig:highlowTSpectra}g-i}), where Nb$_3$Cl$_8$ and Nb$_3$Cl$_7$Br are in the high-temperature $\alpha$-phase, while Nb$_3$Br$_8$ remains in the low-temperature $\beta$-phase. Comparing the spectra of Nb$_3$Cl$_8$ and Nb$_3$Cl$_7$Br between 5 K and 300 K reveals pronounced frequency shifts of $E_u^1$ in Nb$_3$Cl$_8$ and of both $E_u^1$ and $E_u^2$ in Nb$_3$Cl$_7$Br. As discussed below, these shifts provide sensitive signatures of the structural and magnetic phase transitions. The overall evolution of the phonon spectra with Br concentration at 300 K closely resembles that at 5 K: a weak $E_u^2$ mode emerges in Nb$_3$Cl$_7$Br, becomes stronger and shifts to lower frequency in Nb$_3$Br$_8$, and all observed $E_u$ modes soften with increasing Br content. The phonon frequencies at 5 K and 300 K are summarized in Table~\ref{tab:Eu_modes}.

\begin{figure*}[t]
    \centering
    \includegraphics[
        trim=0cm 0cm 0cm 0cm,
        clip,
        width=0.9\textwidth
    ]{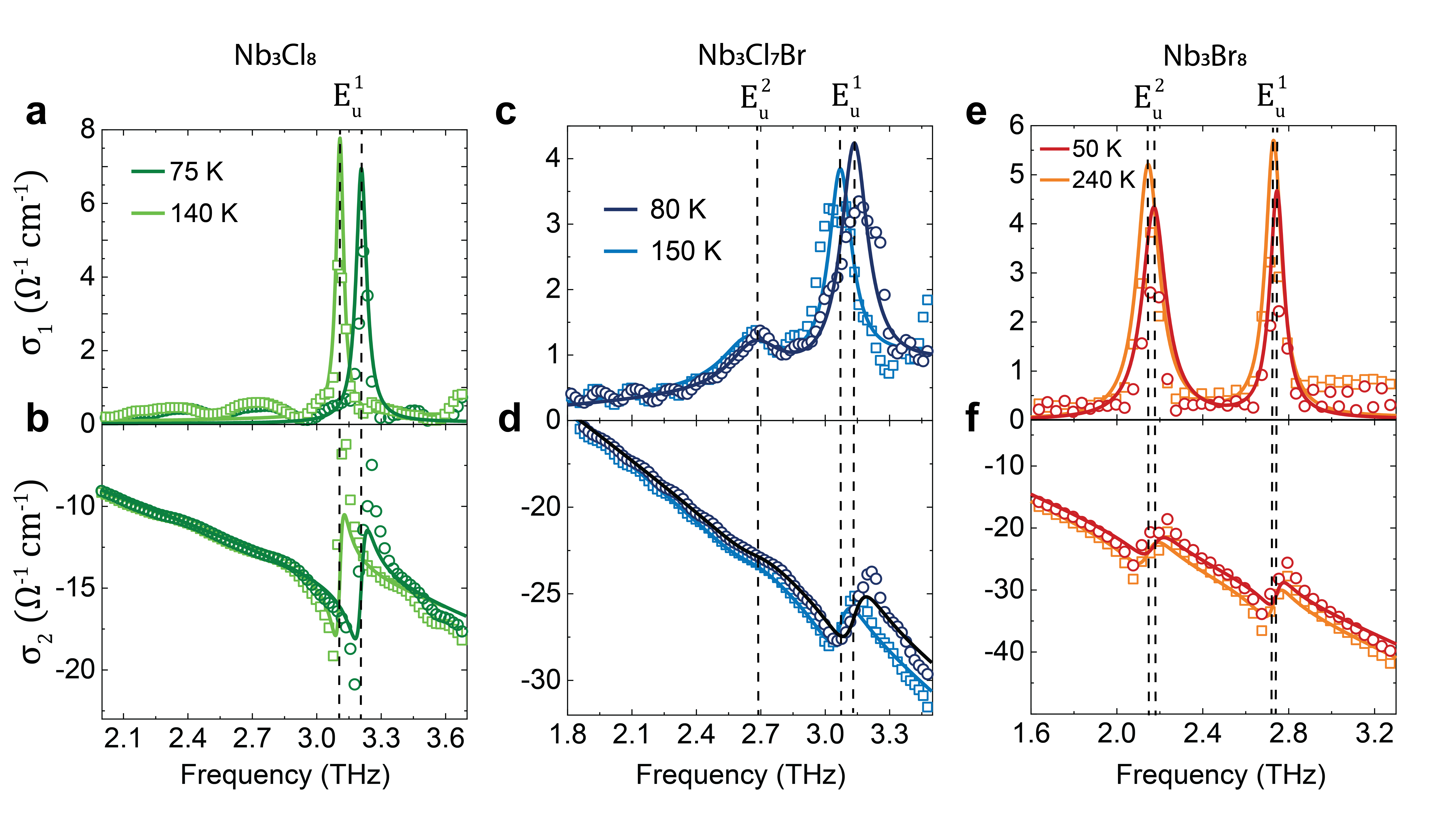}
    \caption{\textbf{Optical conductivity fitted with a Lorentz-oscillator model.}
    (a-f) Real part ($\sigma_1$) and imaginary part ($\sigma_2$) of  optical conductivity  at selected temperatures in Nb$_3$Cl$_8$ (a-b), Nb$_3$Cl$_7$Br (c-d) and Nb$_3$Br$_8$ (e-f). Solid lines are the best simultaneous fit to $\sigma_1$ and $\sigma_2$ using a Lorentz-oscillator model. Clear phonon frequency shifts can be observed across phase transitions in Nb$_3$Cl$_8$ and Nb$_3$Cl$_7$Br.
    }
    \label{fig:conductivity}
\end{figure*}

To track the temperature evolution of phonons and elucidate their relationship to phase transitions, we performed temperature-dependent TDTS measurements on Nb$_3$Cl$_8$, Nb$_3$Cl$_7$Br, and Nb$_3$Br$_8$ from 5 to 300 K. Spectra acquired at selected temperatures during warming are shown in Figs.~\textbf{\ref{fig:TdepSpectra}a-c}, while the corresponding cooling data are presented in Fig.~\textbf{S1}. In Nb$_3$Cl$_8$ (Fig.~\textbf{\ref{fig:TdepSpectra}a}), upon warming from the base temperature, an additional absorption peak appears on the low-frequency side of $E_u^1$ at approximately 110 K. As the temperature increases further, the new peak gains spectral weight while the original $E_u^1$ mode gradually weakens and becomes undetectable above approximately 130 K. During cooling (Fig.~\textbf{S1a}), the reverse evolution occurs between 90 K and 70 K. In contrast, the frequency of $E_u^3$ decreases continuously with increasing temperature without exhibiting peak splitting. The thermal hysteresis of $E_u^1$ together with the simultaneous presence of two $E_u^1$ modes over a finite temperature interval indicates macroscopic coexistence of the $\alpha$- and $\beta$-phases, a characteristic feature of clean first-order transitions that has recently been reported in Nb$_3$Cl$_8$~\cite{huang_suppression_2026}.  Note that the two $E_u^1$ peaks do not result from mode splitting due to symmetry reduction, as only one peak survives deep inside each phase.  

In contrast, Nb$_3$Cl$_7$Br (Fig.~\textbf{\ref{fig:TdepSpectra}b}) exhibits no resolvable coexistence of two phonon branches, as also confirmed by measurements on a thinner sample (Figs.~\textbf{S2a},~\textbf{b}). Instead, pronounced, but broadened frequency shifts are observed for both $E_u^1$ and the newly emerged $E_u^2$ mode. Comparison between the warming and cooling data (Figs.~\textbf{\ref{fig:TdepSpectra}b} and \textbf{S1b}) reveals a clear thermal hysteresis, showing that the transformation remains kinetically irreversible on the experimental timescale, even though its macroscopic spectroscopic discontinuity is strongly rounded.  For Nb$_3$Br$_8$, whose transition temperature lies well above room temperature, the sample remains in the $\beta$-phase throughout the measured temperature range, and only gradual frequency shifts are observed (Fig.~\textbf{\ref{fig:TdepSpectra}c}). The contrast between Nb$_3$Cl$_8$ and Nb$_3$Cl$_7$Br suggests that substitutional disorder fragments the transition and suppresses macroscopic phase coexistence, consistent with theoretical expectations for disorder-modified first-order transitions in low-dimensional systems~\cite{imry_influence_1979, aizenman_rounding_1989}.

From the complex transmission coefficient $T_{xx}$, we extracted the complex optical conductivity of Nb$_3$Cl$_8$, Nb$_3$Cl$_7$Br, and Nb$_3$Br$_8$ using Eqs.~(1)–(4). The real and imaginary parts, $\sigma_1$ and $\sigma_2$, at selected temperatures are shown in Fig.~\ref{fig:conductivity}. As expected, $\sigma_1$ exhibits resonant absorption peaks at the phonon frequencies and is otherwise relatively featureless, whereas $\sigma_2$ displays a characteristic dispersive response, with rapid variations across the resonances superimposed on an approximately linear background arising from higher-energy electronic excitations outside the measured frequency window. This behavior reflects the Kramers–Kronig-related reactive response associated with the phonon absorption peaks in $\sigma_1$.

\begin{figure*}[t]
    \centering
    \includegraphics[
        trim=0cm 0cm 0cm 0cm,
        clip,
        width=0.9\textwidth
    ]{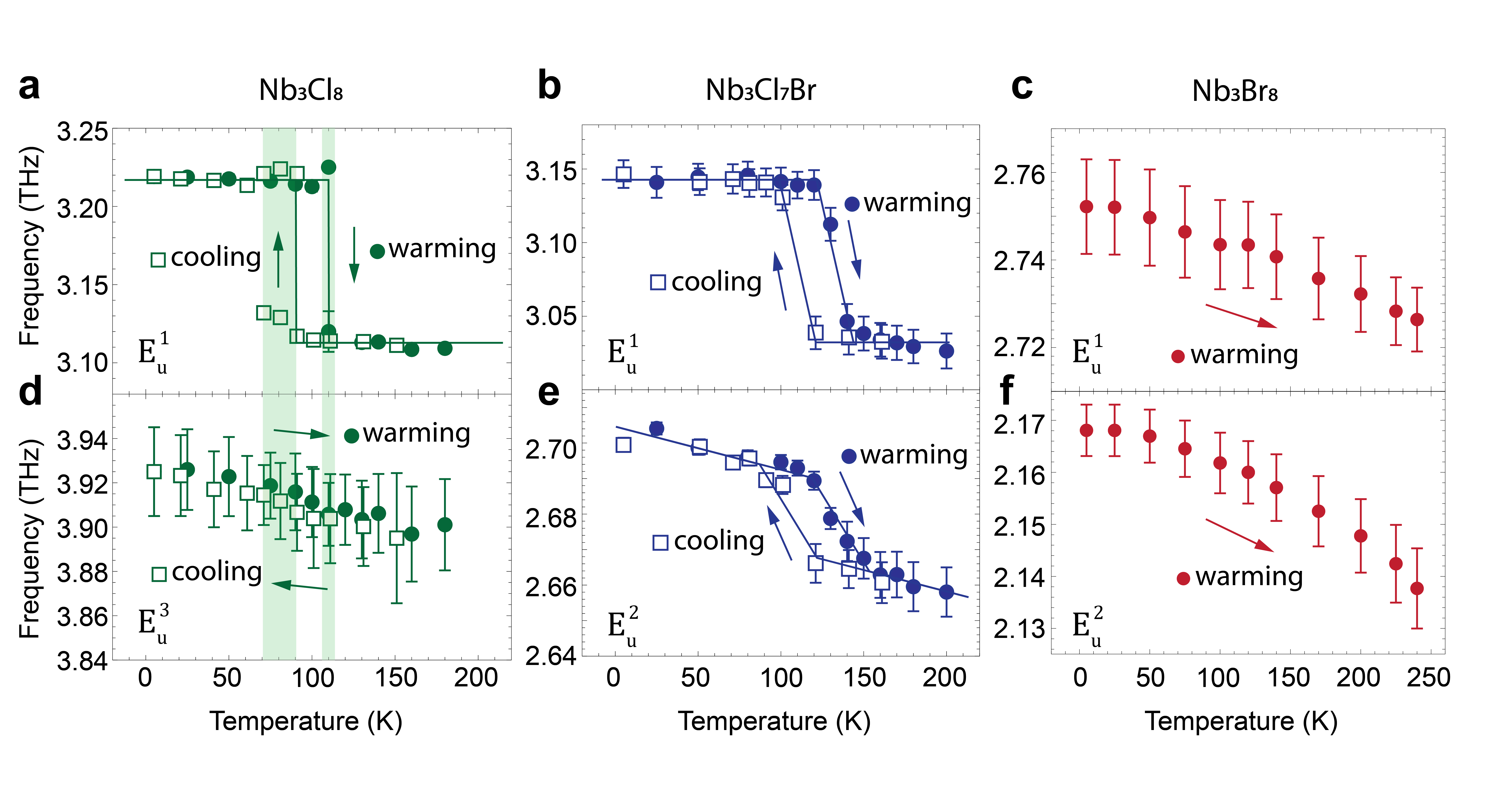}
    \caption{\textbf{Hysteresis in $T$-dependent phonon frequency reflects first-order phase transitions.}
    (a-c) Frequencies of $E_u^1$ phonon as a function of temperature in Nb$_3$Cl$_8$ (a), Nb$_3$Cl$_7$Br (b) and Nb$_3$Br$_8$ (c). (d) Frequencies of $E_u^3$ phonon as a function of temperature in Nb$_3$Cl$_8$. (e-f) Frequencies of $E_u^2$ phonon as a function of temperature in Nb$_3$Cl$_7$Br (e) and Nb$_3$Br$_8$ (f). Solid circles denote warming data, and open squares denote cooling data. Temperatures with coexistence of $\alpha$- and $\beta$-phases in Nb$_3$Cl$_8$ are shaded in green. Error bars represent fitting uncertainties. Solid lines are guide to the eyes. Note the very different frequency scales in the individual panels.
    }
    \label{fig:phononfreq}
\end{figure*}

The optical conductivity spectra are well described by a sum of Lorentz oscillators. The real and imaginary parts, $\sigma_1$ and $\sigma_2$, were fitted simultaneously using

\begin{equation}
\sigma_1(\omega) = A \omega^2+ \sum_i C_i \frac{\omega^2/\tau_i}{(\omega_i^2-\omega^2)^2+\omega^2/\tau_i^2},
\end{equation}

\begin{equation}
\sigma_2(\omega) = B \omega - \sum_i C_i \frac{\omega(\omega_i^2-\omega^2)}{(\omega_i^2-\omega^2)^2+\omega^2/\tau_i^2},
\end{equation}

\noindent where $\omega_i$ and $\tau_i$ denote the resonance frequency and lifetime of the $i$th oscillator. The coefficients $C_i$ characterize the oscillator strengths, while the terms proportional to $A\omega^2$ and $B\omega$ account for the background arising from higher-energy excitations outside the measured frequency range. The best-fit curves are overlaid with the experimental data in Fig.~\ref{fig:conductivity}. Small discrepancies near the resonance frequencies likely originate from uncertainties in the phase of the transmitted THz signal when the transmission amplitude becomes very small.

The phonon frequencies extracted from the Lorentz fits are summarized in Fig.~\ref{fig:phononfreq}. In Nb$_3$Cl$_8$, $E_u^1$ exhibits abrupt frequency shifts of approximately 0.1 THz, occurring near 110 K during warming and 90 K during cooling, whereas the frequency of $E_u^3$ evolves only weakly and continuously with temperature. Coexistence of the $\alpha$- and $\beta$- phases is observed over a finite temperature interval, occurring near 110 K during warming and between 70 and 90 K during cooling (Figs.~\textbf{\ref{fig:phononfreq}a,d}). In contrast, Nb$_3$Cl$_7$Br exhibits smoother frequency evolution. Both $E_u^1$ and the emergent $E_u^2$ undergo gradual shifts of approximately 0.12 THz and 0.07 THz, respectively, centered near 135 K during warming and 110 K during cooling. No resolvable coexistence of the $\alpha$- and $\beta$-phases is observed but the hysteresis remains present (Figs.~\textbf{\ref{fig:phononfreq}b,e}). Measurements on a thinner Nb$_3$Cl$_7$Br sample (thickness $\sim$80 $\upmu$m) with finer temperature steps yield similar results (Figs.~\textbf{S2c},\textbf{d}), showing only a single $E_u^1$ peak whose frequency evolves smoothly across the transition. In Nb$_3$Br$_8$, whose transition temperature lies above the measured temperature range, the sample remains in the $\beta$-phase and only weak, continuous phonon softening is observed upon warming (Figs.~\textbf{\ref{fig:phononfreq}c,f}). 

\begin{figure*}[t]
    \centering
    \includegraphics[
        trim=0cm 0cm 0cm 0cm,
        clip,
        width=0.9\textwidth
    ]{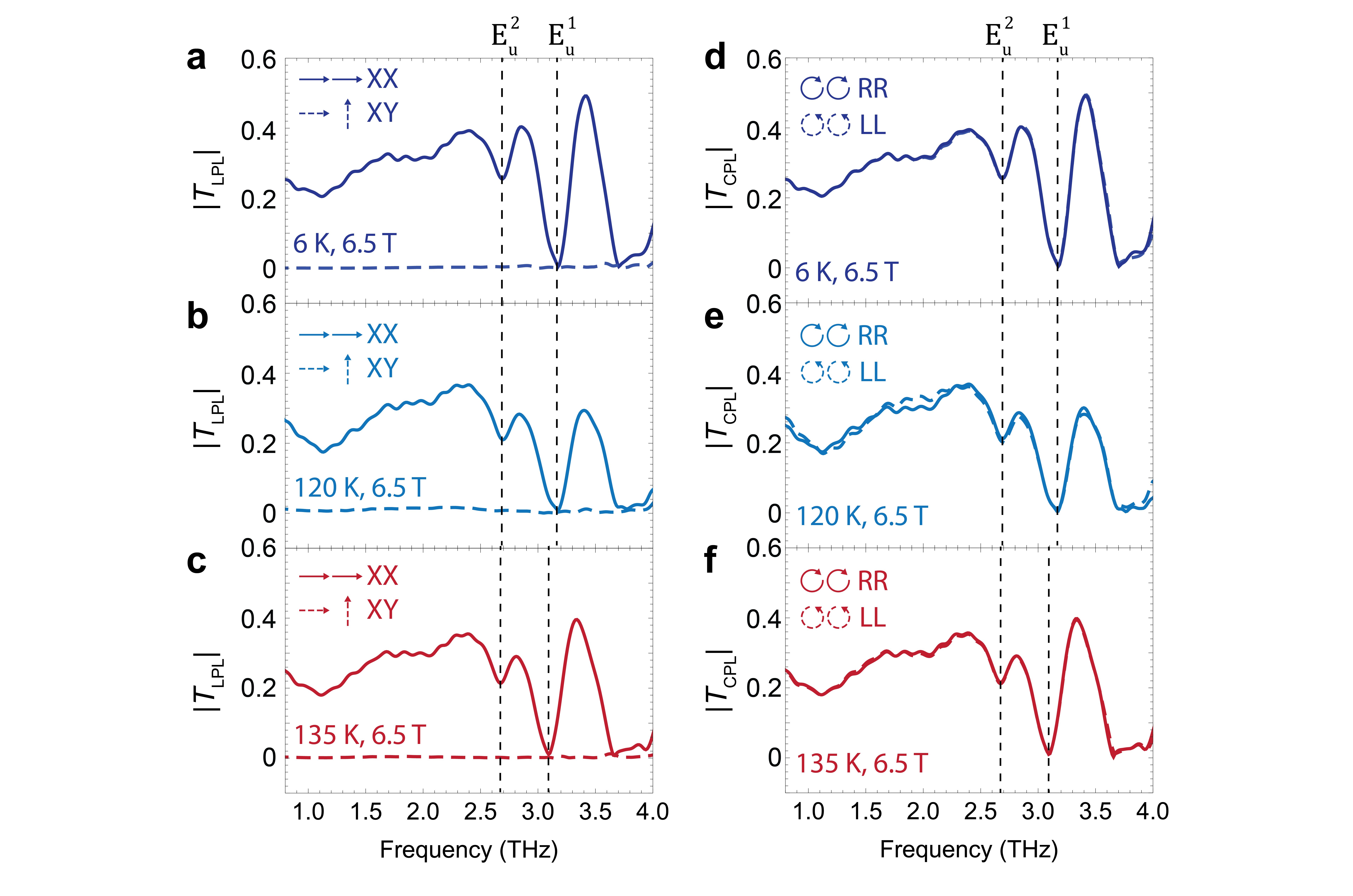}
    \caption{\textbf{Transmission coefficients in linear and circular basis characterize structural and magnetic ground states.}
   (a-f) The magnitudes of the transmission coefficients in linear basis $T_{LPL}$ ($T_{xx}$ and $T_{xy}$) (a-c), and circular basis $T_{CPL}$ ($T_{RR}$ and $T_{LL}$) (d-f) of Nb$_3$Cl$_7$Br as a function of frequency at 6 K, far below the magnetic transition temperature (a and b), 120 K, right below the magnetic transition temperature (b and e) and 135 K, right above the magnetic transition temperature (c and f). The almost-zero $|T_{xy}|$ is consistent with $\bar{3}m$ point group symmetry and the almost-identical spectra of $|T_{RR}|$ and $T_{LL}$ confirms a non-magnetic ground state.
    }
    \label{fig:Bfield}
\end{figure*}

We also extracted the phonon linewidths, $\Gamma=1/2\pi\tau$, from the Lorentz fits, as shown in Fig.~\textbf{S3}. Among the three compounds, Nb$_3$Cl$_7$Br exhibits the largest linewidth for the $E_u^1$ mode, approximately three times that of Nb$_3$Br$_8$ and five times that of Nb$_3$Cl$_8$. This broadening is evident in the transmission spectra (Figs.~\textbf{\ref{fig:highlowTSpectra}e,h} and \textbf{\ref{fig:TdepSpectra}b}) and in the corresponding conductivity spectra (Fig.~\textbf{\ref{fig:conductivity}c}). The substantially larger linewidth in Nb$_3$Cl$_7$Br is consistent with enhanced disorder-induced phonon broadening arising from halogen substitution. As discussed later, this disorder strongly modifies the characteristics of the first-order phase transition.

To further investigate the nature of the $E_u$ phonons and the effect of Br substitution on the ground state, we performed THz polarimetry measurements on Nb$_3$Cl$_7$Br under a magnetic field of 6.5 T. In  Figs.~\textbf{\ref{fig:Bfield}a-c} we present the transmission matrix elements in the linear polarization basis at several representative temperatures: 6 K, well below the transition temperature; 120 K, just below the transition; and 135 K, just above the transition. At all temperatures, nearly all spectral weight resides in the parallel channel $T_{xx}$, while the cross-polarized response $T_{xy}$ remains negligible. Furthermore, no additional phonon modes or splitting of the $E_u$ modes are observed across the phase transition.  These observations are consistent with the preservation of threefold rotational and inversion symmetries across the transition. Within sensitivity of the experimental setup, our data support assignment of the low-temperature structure to the $R\bar{3}m$ space group and disfavor lower-symmetry groups such as $R3$ and $C2/m$.

We transform the transmission matrix into the circular polarization basis and examine the amplitudes of $T_{RR}$ and $T_{LL}$ via the relation $T_{RR,LL} = T_{xx} \pm i T_{xy}$.  As seen in Figs.~\textbf{\ref{fig:Bfield}d-f}, at 6.5 T, the spectra in the two circular channels are nearly identical at all temperatures. The absence of circular dichroism is consistent with a paramagnetic $\alpha$-phase and a nonmagnetic $\beta$-phase, indicating that Br substitution does not alter the nonmagnetic ground state despite substantially modifying the character of the first-order transition. Moreover, the identical $T_{RR}$ and $T_{LL}$ spectra indicate that the infrared-active $E_u$ phonons do not possess observable net angular momentum and behave as conventional phonons, in contrast to the chiral Raman-active $E_g$ phonons recently reported in Nb$_3$X$_8$~\cite{tang_chiral_2025, duan_coexistence_2023}.

\section{Discussion}

Disorder strongly influences the behavior of phase transitions. For second-order transitions, the Harris criterion addresses the stability of the critical behavior against weak quenched disorder. It states that disorder is irrelevant when $d\nu \geq 2$, where $d$ is the spatial dimensionality and $\nu$ is the correlation length critical exponent~\cite{harris_effect_1974}. When this condition is violated, disorder modifies the critical behavior and can lead to a new universality class.

In contrast, first-order phase transitions are characterized by finite correlation lengths, rendering the original Harris criterion inapplicable. Recognizing this distinction, Imry and Wortis developed an analogous argument for first-order transitions~\cite{imry_influence_1979}. They showed that the competition between the bulk free-energy gain and the interfacial energy cost can destabilize macroscopic phase coexistence and fragment the transition into spatially separated domains. This instability becomes particularly pronounced in low-dimensional systems. Building upon these ideas, Aizenman and Wehr demonstrated that arbitrarily weak quenched disorder rounds the thermodynamic discontinuities associated with first-order phase transitions for $d\leq2$~\cite{aizenman_rounding_1989}. As a consequence, sharp first-order transitions and macroscopic phase coexistence are expected to become increasingly fragile in low-dimensional systems in the presence of disorder.   Related work on the random field Ising model (RFIM) illuminated nonequilibrium aspects of this and showed how above a critical level of disorder discontinuities are smeared out, while hysteresis and memory-dependent switching may survive~\cite{sethna1993hysteresis}.

Our observations in Nb$_3$Cl$_{8-x}$Br$_x$ provide a platform for experimentally examining these theoretical predictions. In the intermediate compound Nb$_3$Cl$_7$Br, the phonon linewidths are substantially larger than those of the end compounds Nb$_3$Cl$_8$ and Nb$_3$Br$_8$, providing spectroscopic evidence for enhanced disorder arising from Br substitution. In Nb$_3$Cl$_8$, the observation of two $E_u^1$ peaks across the transition indicates macroscopic coexistence of the $\alpha$- and $\beta$-phases, a hallmark of a clean first-order transition. Near the transition temperature, the free energies of the two phases become nearly degenerate, allowing domains of both phases to coexist. Upon cooling, the $\beta$-phase becomes energetically favorable, and the bulk free-energy gain associated with the low-temperature phase eventually overcomes the interfacial energy cost, leading to the growth of the $\beta$-phase at the expense of the $\alpha$-phase.

In Nb$_3$Cl$_7$Br, although the system is not strictly two dimensional (2D), its quasi-2D nature together with substantial substitutional disorder favors the fragmentation of the emergent phases and leads to a broadened first-order transition~\cite{pasco_tunable_2019}. Compared with Nb$_3$Cl$_8$, the transition remains hysteretic but exhibits a much smoother evolution of the phonon frequencies and no resolvable macroscopic phase coexistence. The persistence of hysteresis suggests that the transition becomes a disorder-rounded descendant of a clean first-order transition.  Disorder-induced spatial inhomogeneity and pinning cause different regions of the sample to transform at different temperatures, giving rise to a pronounced history dependence. At the same time, disorder suppresses macroscopic phase coexistence by breaking the system into many small, locally favored domains rather than two well-defined bulk phases separated by stable interfaces.  These observations are qualitatively consistent with the Imry-Wortis picture and motivated by the Aizenman-Wehr scenario, in which quenched disorder in low-dimensional systems destabilizes macroscopic phase coexistence and rounds the thermodynamic discontinuities associated with first-order transitions.  The occurrence of hysteresis without coexistence is not contradictory~\cite{sethna1993hysteresis}.  The energetic barrier that enables  macroscopic coexistence over a finite temperature interval in a clean sample is a nucleation barrier, set by the competition between bulk free-energy gain and interfacial cost, and it shrinks as the relevant length scale shrinks. The barrier responsible for hysteresis in the disordered sample is instead a depinning barrier, set by the strength of the local disorder potential itself and largely independent of domain size. Disorder can therefore simultaneously suppress the nucleation-limited, macroscopic coexistence signature while sustaining or even enhancing a depinning-limited hysteretic response, since both effects share a common origin in the quenched disorder but act on the two observables in opposite ways.

Our observations resemble several experimental results that have been interpreted in the context of the RFIM~\cite{sethna1993hysteresis}.  In capillary condensation of 4He in silica aerogel~\cite{Bonnet2008Aerogel}, the aerogel provides quenched random local preferences for liquid and vapor, while pressure plays the role of the uniform field. Depending on aerogel microstructure and temperature, condensation changes from an abrupt, avalanche-like event to a smooth filling curve, yet adsorption–desorption hysteresis remains. This is well described as a disorder-driven transition in metastable dynamics rather than ordinary equilibrium liquid–vapor coexistence.  Berger et al.~\cite{Berger2000HysteresisCriticality} studied Co/CoO bilayers in which structural disorder could be tuned. They measured magnetic hysteresis loops as a function of applied magnetic field and found a disorder-driven change from loops containing a sharp magnetization reversal to smooth loops, with scaling behavior near an apparent critical disorder. This was interpreted as experimental evidence for the nonequilibrium critical point predicted by the zero-temperature random-field Ising model.  Below the critical disorder, reversal occurs through a macroscopic avalanche, but above it, reversal is distributed among many smaller avalanches, while hysteresis remains in both regimes.  These RFIM examples should be taken as a conceptual analogy for the survival of hysteresis after rounding, as we have not demonstrated a precise mapping of the RFIM to the present case.

In all of this, it is important to note that the bulk Nb$_3$Cl$_{8-x}$Br$_x$ crystals studied here are quasi-2D rather than strictly 2D. Nevertheless, the observed disorder-induced broadening of the first-order transition already suggests that reduced dimensionality plays an important role. Although experimentally challenging, it would be highly desirable to exfoliate the van der Waals Nb$_3$Cl$_{8-x}$Br$_x$ compounds to the few-layer or monolayer limit to realize a truly 2D system and directly investigate the interplay between disorder and first-order phase transitions, paired with advanced imaging techniques~\cite{vidas_imaging_2018, mattoni_striped_2016}. A dimensional crossover in the critical behavior and the character of the transition is expected as the sample thickness is reduced.

\section{Summary}
In summary, we investigated the effect of disorder on first-order phase transitions in Nb$_3$Cl$_{8-x}$Br$_x$ ($x=0$, 1, and 8) using TDTS to probe the lattice dynamics, including a new phonon branch activated by Br-substitution. We find that Nb$_3$Cl$_8$ exhibits a clean first-order transition, manifested by macroscopic coexistence of the $\alpha$- and $\beta$-phases, and a relatively narrow thermal hysteresis. In contrast, Nb$_3$Cl$_7$Br exhibits substantially broadened phonon linewidths, providing spectroscopic evidence for enhanced substitutional disorder introduced by Br incorporation. The transition in Nb$_3$Cl$_7$Br remains hysteretic but displays a much smoother evolution of both the original $E_u^1$ and the Br-activated $E_u^2$ phonons, with no resolvable macroscopic phase coexistence. These observations are consistent with disorder-induced fragmentation of the transition into locally favored domains and the suppression of macroscopic phase coexistence.  These results are qualitatively consistent with the Imry–Wortis picture and with the disorder-rounding physics established by Aizenman and Wehr.  Our results show that disorder suppresses resolvable macroscopic two-phase coexistence, while preserving thermal hysteresis over experimentally accessible times.  In a disorder-rounded scenario, the hysteresis is expected to depend on the measurement timescale and to vanish in the ideal equilibrium limit.  Measurements of the time dependence, local domain structure, and thickness evolution would provide useful future tests of this interpretation.  More broadly, van der Waals materials offer a promising avenue for exploring disorder-mediated first-order transitions approaching the 2D limit.

\bigskip

\noindent\textbf{Acknowledgments:}  THz experiments were supported by the US Department of Energy DE-SC0025245 ``Dynamics and time-evolution in quantum magnets as probed by new nonlinear THz spectroscopies.” Instrumentation development at JHU, which made these measurements possible, was supported by the Gordon and Betty Moore Foundation EPiQS Initiative Grant GBMF-9454 to NPA. Materials development was supported by the U.S. Department of Energy, Office of Basic Energy Sciences, Division of Material Sciences and Engineering under Award No. DE-SC0019331 to Institute for Quantum Matter at JHU.

\bibliographystyle{unsrt}
\bibliography{reference_cleaned}

\end{document}


\title{Supplementary Materials for\\
Hysteresis without coexistence: disorder-rounded first-order transitions in a van der Waals magnet}

\author{Xiaoyu Guo}
\affiliation{William H. Miller III Department of Physics and Astronomy, The Johns Hopkins University, Baltimore, Maryland 21218, USA}

\author{Abby N. Neill}
\affiliation{Department of Chemistry,
The Johns Hopkins University, Baltimore, Maryland 21218, USA}

\author{Christopher M. Pasco}
\affiliation{Department of Chemistry,
The Johns Hopkins University, Baltimore, Maryland 21218, USA}

\author{Tyrel M. McQueen}
\affiliation{William H. Miller III Department of Physics and Astronomy, The Johns Hopkins University, Baltimore, Maryland 21218, USA}
\affiliation{Department of Chemistry,
The Johns Hopkins University, Baltimore, Maryland 21218, USA}
\affiliation{Department of Materials Science and Engineering,
The Johns Hopkins University, Baltimore, Maryland 21218, USA}

\author{N. P. Armitage}
\affiliation{William H. Miller III Department of Physics and Astronomy, The Johns Hopkins University, Baltimore, Maryland 21218, USA}

\date{\today}

\begin{abstract}  

\end{abstract}

\maketitle
\setcounter{section}{0}
\renewcommand{\thesection}{S\arabic{section}}

\setcounter{figure}{0}
\renewcommand{\thefigure}{S\arabic{figure}}

\setcounter{table}{0}
\renewcommand{\thetable}{S\arabic{table}}

\setcounter{equation}{0}
\renewcommand{\theequation}{S\arabic{equation}}

\section{Terahertz transmission spectra during cooling}
\begin{figure}[h]
    \centering
    \includegraphics[
        trim=0cm 0cm 0cm 0cm,
        clip,
        width=\textwidth
    ]{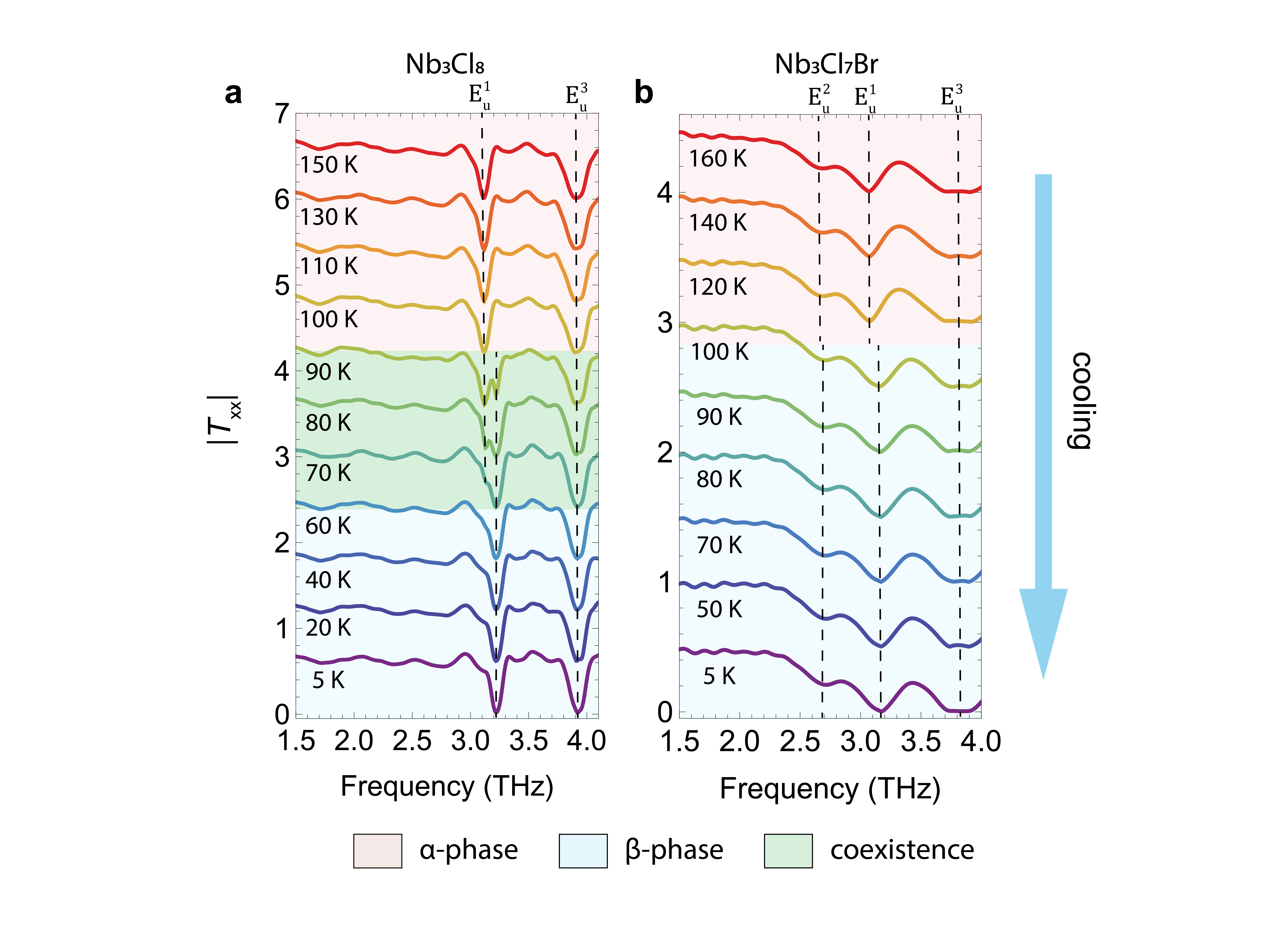}
    \caption{\textbf{$T$-dependent transmission spectra measured during cooling down.}
    (a-c) The magnitude of the transmission coefficient $T_{xx}$ with temperature in Nb$_3$Cl$_8$ (a) and Nb$_3$Cl$_7$Br (b). Measurements were taken during cooling down. The temperature range where they are in the $\alpha$-and $\beta$-phases and their coexistence are shaded in red, blue and green, respectively.
}
    \label{fig:coolSpectra}
\end{figure}

\section{Terahertz transmission spectra measured from a thinner Nb$_3$Cl$_7$Br}

\begin{figure}[h]
    \centering
    \includegraphics[
        trim=0cm 2cm 0cm 0cm,
        clip,
        width=\textwidth
    ]{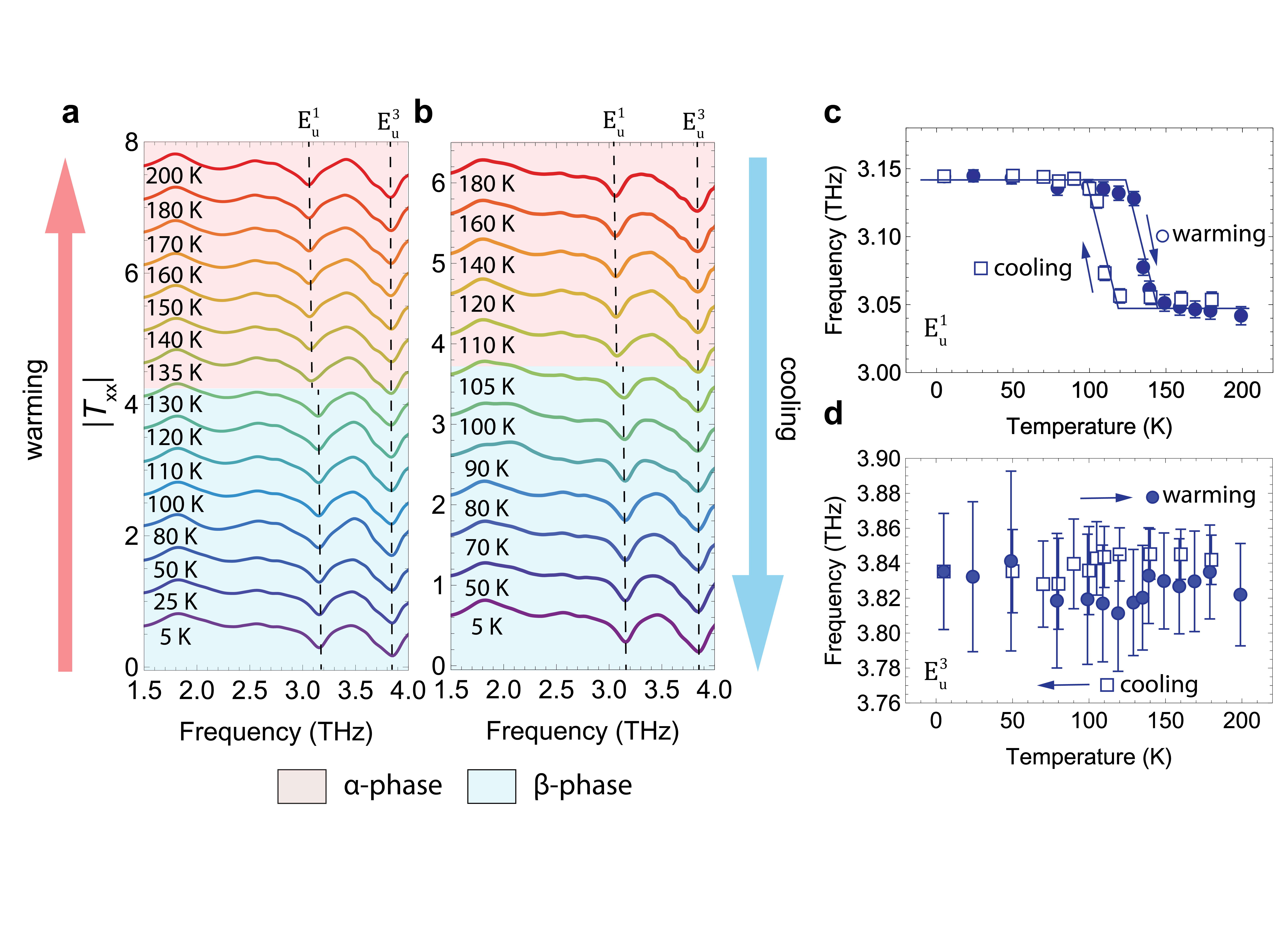}
    \caption{\textbf{$T$-dependent transmission spectra measured from a thinner Nb$_3$Cl$_7$Br sample.}
    (a and b) The magnitude of the transmission coefficient $T_{xx}$ with temperature measured during warming up (a) and cooling down (b). (c and d) Frequencies of $E_u^1$ (c) and $E_u^3$ (d) with temperature. Solid circles are data from warming up and white squares are data from cooling down. Error bars represent fitting uncertainties. Solid lines are guide to the eyes.
}
    \label{fig:NCB2}
\end{figure}

\newpage
\section{Phonon linewidth from Lorentz model fitting}
\begin{figure}[h]
    \centering
    \includegraphics[
        trim=0cm 0cm 0cm 0cm,
        clip,
        width=\textwidth
    ]{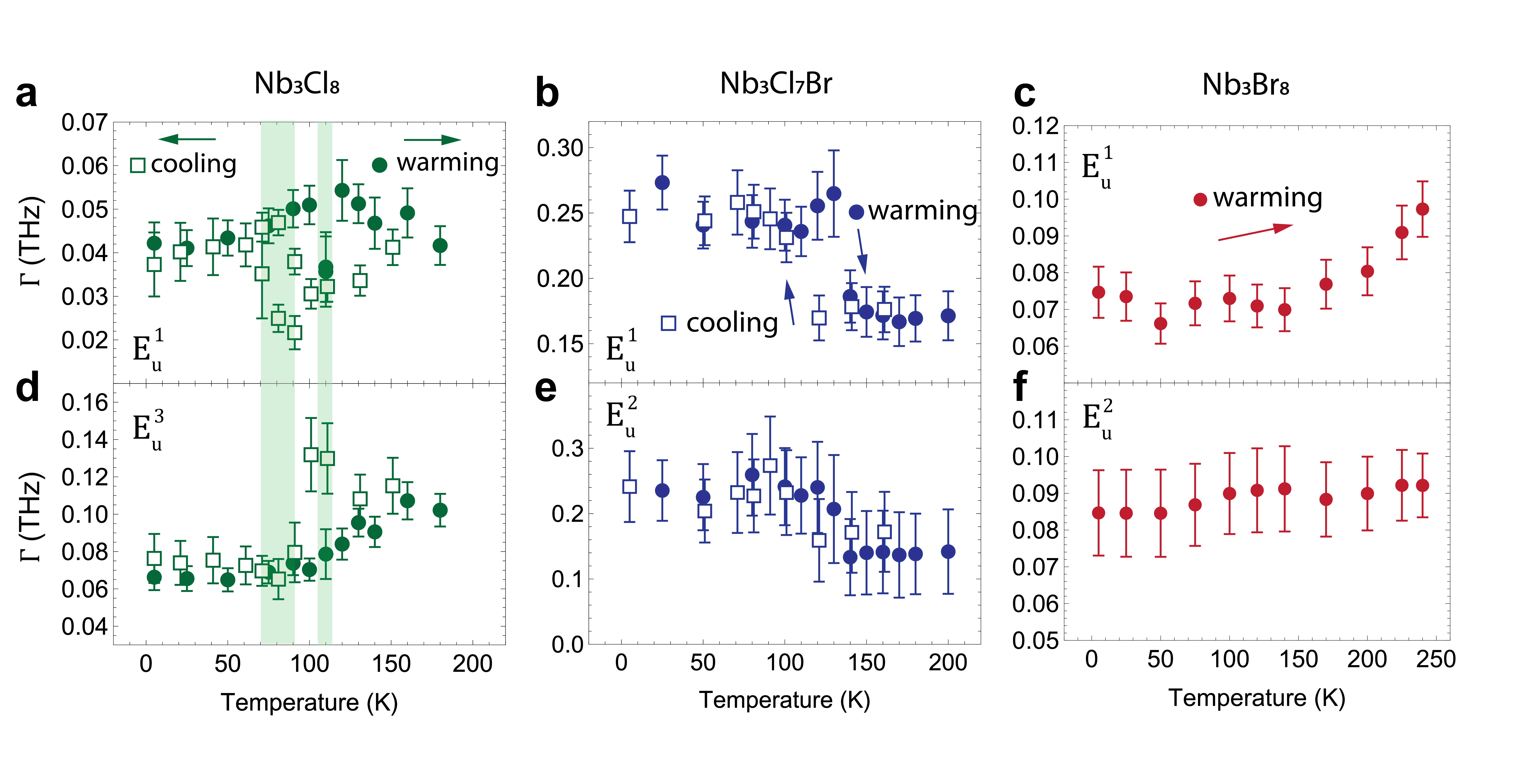}
    \caption{\textbf{Scattering rates obtained from Lorentz model fitting.}
    (a-c) Scattering rate of $E_u^1$ phonon with temperature in Nb$_3$Cl$_8$ (a), Nb$_3$Cl$_7$Br (b) and Nb$_3$Br$_8$ (c). (d) Scattering rate of $E_u^3$ phonon with temperature in Nb$_3$Cl$_8$. (e and f) Scattering rate of $E_u^2$ phonon with temperature in Nb$_3$Cl$_7$Br (e) and Nb$_3$Br$_8$ (f). Solid circles are data from warming up and white squares are data from cooling down. Temperatures with coexistence of $\alpha$- and $\beta$-phases in Nb$_3$Cl$_8$ are shaded in green. Error bars are fitting errors.
}
    \label{fig:scatteringRate}
\end{figure}